\documentclass{styles/svproc}
\usepackage{url}
\usepackage{graphicx}
\usepackage{float}

\begin{document}
\mainmatter              % start of a contribution

\title{Advanced System Engineering Approaches to Emerging Challenges in Planetary and Deep-Space Exploration}

\titlerunning{Advanced System Engineering for Planetary Exploration}

\author{J. de Curt\`o\inst{1,2,10} \and Cristina LiCalzi\inst{1,3} \and Julien Tubiana Warin\inst{1,4} \and Jack Gehlert\inst{1,5} \and Brian Langbein\inst{1,3} \and Alexandre Gamboa \inst{1,6} \and Chris Sixbey\inst{1,3} \and William Maguire\inst{1,3} \and Santiago Fernández\inst{1} \and Álvaro Maestroarena\inst{1} \and Alex Brenchley\inst{1} \and Logan Maroclo\inst{1,3} \and Philemon Mercado\inst{1,3} \and Joshua DeJohn\inst{1,7} \and Cesar Velez\inst{1,8} \and Ethan Dahmus\inst{1,3} \and Taylor Steinys\inst{1,7} \and David Fritz \inst{1,3} \and I. de Zarz\`a\inst{9,10}}

\authorrunning{J. de Curt\`o et al.}

\tocauthor{J. de Curt\`o et al.}

\institute{Escuela Técnica Superior de Ingeniería (ICAI), Universidad Pontificia Comillas, 28015 Madrid, Spain
\and Department of Computer Applications in Science \& Engineering, BARCELONA Supercomputing Center, 08034 Barcelona, Spain
\and Dept. of Mechanical Engineering, University of Maryland, College Park, USA
\and School of Engineering, EPFL, 1015 Lausanne, Switzerland
\and College of Engineering, University of Michigan, Ann Arbor, USA
\and Grainger College of Engineering, University of Illinois at Urbana-Champaign, USA
\and Dept. of Mechanical Engineering, Iowa State University, USA
\and Dept. of Mechanical Engineering, University of Wisconsin-Madison, USA
\and Departamento de Informática e Ingeniería de Sistemas, Universidad de Zaragoza, 50009 Zaragoza, Spain
\and Estudis d'Informàtica, Multimèdia i Telecomunicació, Universitat Oberta de Catalunya, 08018 Barcelona, Spain}

\maketitle              % typeset the title of the contribution

\begin{abstract}
This paper presents innovative solutions to critical challenges in planetary and deep-space exploration electronics. We synthesize findings across diverse mission profiles, highlighting advances in: (1) MARTIAN positioning systems with dual-frequency transmission to achieve $\pm$1m horizontal accuracy; (2) artificial reef platforms for Titan's hydrocarbon seas utilizing specialized sensor arrays and multi-stage communication chains; (3) precision orbital rendezvous techniques demonstrating novel thermal protection solutions; (4) miniaturized CubeSat architectures for asteroid exploration with optimized power-to-mass ratios; and (5) next-generation power management systems for MARS rovers addressing dust accumulation challenges. These innovations represent promising directions for future space exploration technologies, particularly in environments where traditional Earth-based electronic solutions prove inadequate. The interdisciplinary nature of these developments highlights the critical intersection of aerospace engineering, electrical engineering, and planetary science in advancing human exploration capabilities beyond Earth orbit.
\end{abstract}

\section{Introduction}
\label{sn:introduction}
The exploration of our solar system presents unprecedented challenges for electronic systems that must operate reliably in environments far more extreme than those encountered on Earth. As humanity's ambitions extend beyond Earth orbit to Mars \cite{vasavada2014,grotzinger2015}, asteroids, and the outer planets, aerospace electronics must evolve to meet these new frontiers' unique demands. This paper examines innovative solutions developed through a series of aerospace electronics projects addressing critical challenges in interplanetary navigation, power management, communication \cite{koktas2024}, and scientific instrumentation.

\begin{figure}[H]
    \centering
    \includegraphics[width=0.9\textwidth]{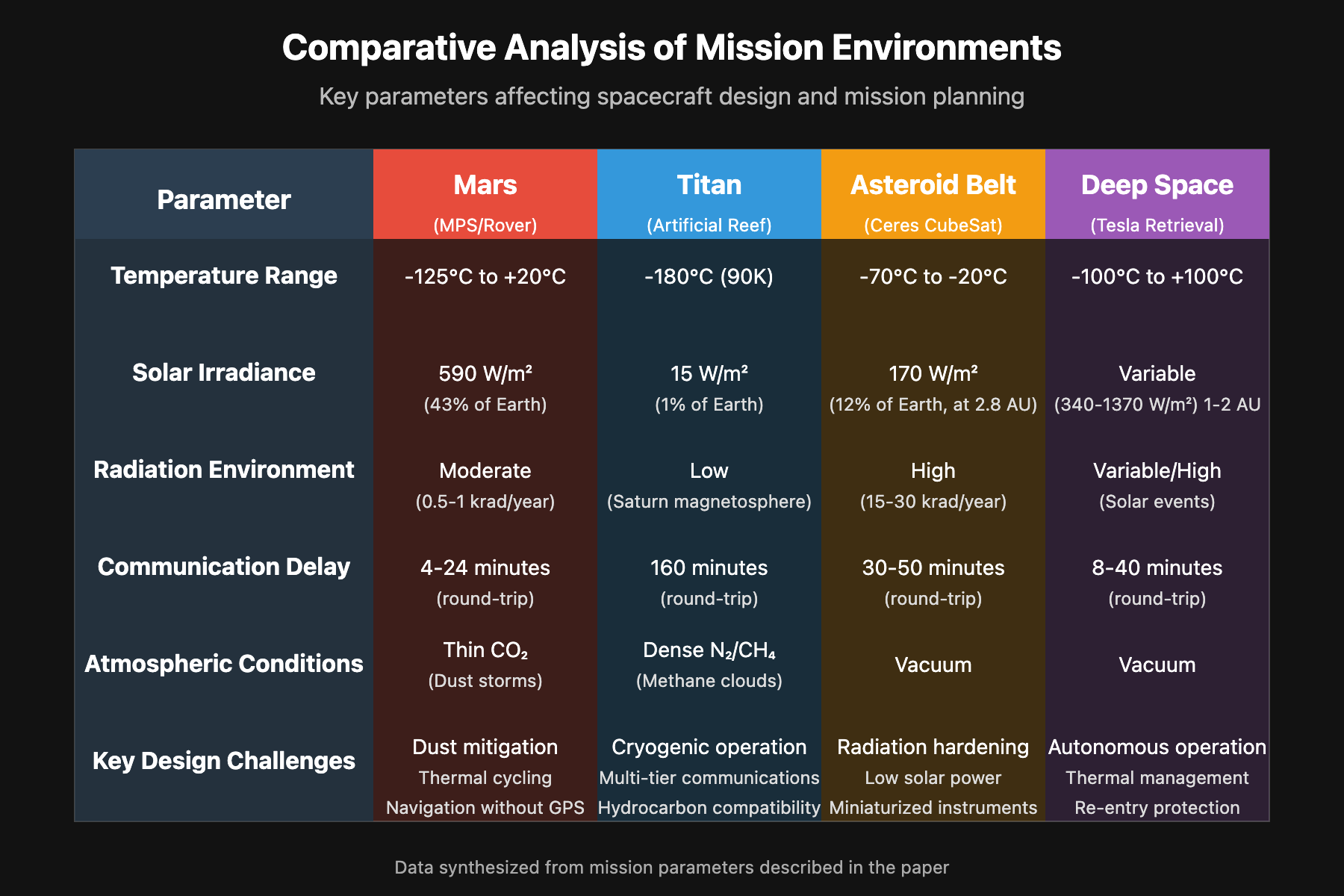}
    \caption{Comparative analysis of mission environments across the solar system destinations discussed in this paper. The chart illustrates key environmental parameters including temperature range, solar irradiance, radiation environment, communication delay, atmospheric conditions, and resulting design challenges. Each destination presents unique challenges: Mars with its dust storms and temperature extremes; Titan with its cryogenic (90K) temperatures and hydrocarbon seas; the Asteroid Belt with high radiation and low solar flux; and Deep Space with variable conditions and re-entry challenges. These environmental differences drive the specific design solutions developed for each mission profile.}
    \label{fgr:mission-environments}
\end{figure}

Recent successful missions, including NASA's Perseverance rover, China's Tianwen-1 \cite{chen2024_2}, and ESA's ExoMars \cite{gonzalez2024}, demonstrate the increasing global interest in planetary exploration \cite{barth2024,brinkmann2025}. However, these missions reveal persistent challenges in several domains: power generation and distribution in dusty, low-irradiance environments; reliable communication across vast distances; precise navigation without GPS infrastructure; and instrumentation capable of withstanding extreme temperatures and radiation \cite{deCurto2023_2}. The projects presented herein directly address these challenges through novel applications of existing technologies and development of new approaches.

The six projects synthesized in this article represent diverse mission profiles spanning multiple planetary bodies. On Mars \cite{deCurto2024}, the proposed MARTIAN Positioning System addresses the critical need for precise navigation infrastructure to support future robotic and human missions. For power management challenges in the Martian environment, advanced Power Control and Distribution Unit (PCDU) systems \cite{xintong2024} demonstrate adaptive solutions to dust accumulation and seasonal variation in solar irradiance. In deep space, innovative approaches to orbital rendezvous \cite{deCurto2024_2} and retrieval operations showcase advances in autonomous navigation and thermal protection systems necessary for complex orbital maneuvers. For asteroid exploration, CubeSat architectures \cite{cannon2025,villela2019} with optimized instruments demonstrate cost-effective scientific return from small bodies. The High-Speed Observational Satellite (HSO-SAT) system presents novel approaches to global surveillance with applications for both Earth and other planetary bodies. Finally, ambitious proposals for TITAN exploration \cite{goossens2024}, including artificial reef \cite{baine2001} structures designed for astrobiological investigation, push the boundaries of electronics designed to function in cryogenic hydrocarbon environments.

Common to these diverse missions is the need for radiation-hardened electronics, efficient power management, thermal control in extreme environments, reliable communications across vast distances, and fault-tolerant systems that can operate with minimal human intervention. Each project demonstrates unique approaches to these shared challenges, contributing valuable insights to the field of aerospace electronics.

As shown in Figure~\ref{fgr:mission-environments}, the diverse environmental conditions across solar system destinations necessitate tailored approaches to power generation, thermal management, and communication architectures.

The following sections detail the technological innovations in: (\ref{sn:positioning}) positioning and navigation systems for non-Earth environments; (\ref{sn:powergeneration}) power generation and distribution strategies for challenging planetary conditions; (\ref{sn:communicationdeepspace}) communication architectures for deep space missions; (\ref{sn:thermalmanagement}) thermal management solutions for extreme environments;  
(\ref{sn:autonomousoperation}) novel approaches to autonomous operation and fault tolerance;
 and (\ref{sn:simulationframework}) simulation framework. By examining these innovations collectively, identify emerging trends and promising directions for future development in aerospace electronics in Section (\ref{sn:conclusion}).

\section{Positioning and Navigation Systems for Non-Earth Environments}
\label{sn:positioning}

Precise positioning and navigation capabilities are foundational for successful exploration of other planetary bodies. While Earth benefits from the Global Positioning System (GPS) and other GNSS constellations, no such infrastructure exists beyond our planet.

The proposed Martian Positioning System represents a significant advancement in planetary navigation infrastructure. Unlike Earth-based navigation systems, the MARTIAN Positioning System must contend with unique Martian environmental challenges while providing reliable positioning data for future robotic and human missions. The system architecture consists of 24 satellites distributed across six orbital planes, providing global coverage with position accuracy of $\pm$1m horizontally and $\pm$2m vertically--specifications that would enable precise scientific measurements and safe autonomous vehicle operation.

\begin{figure}[htbp]
    \centering
    \includegraphics[width=0.9\textwidth]{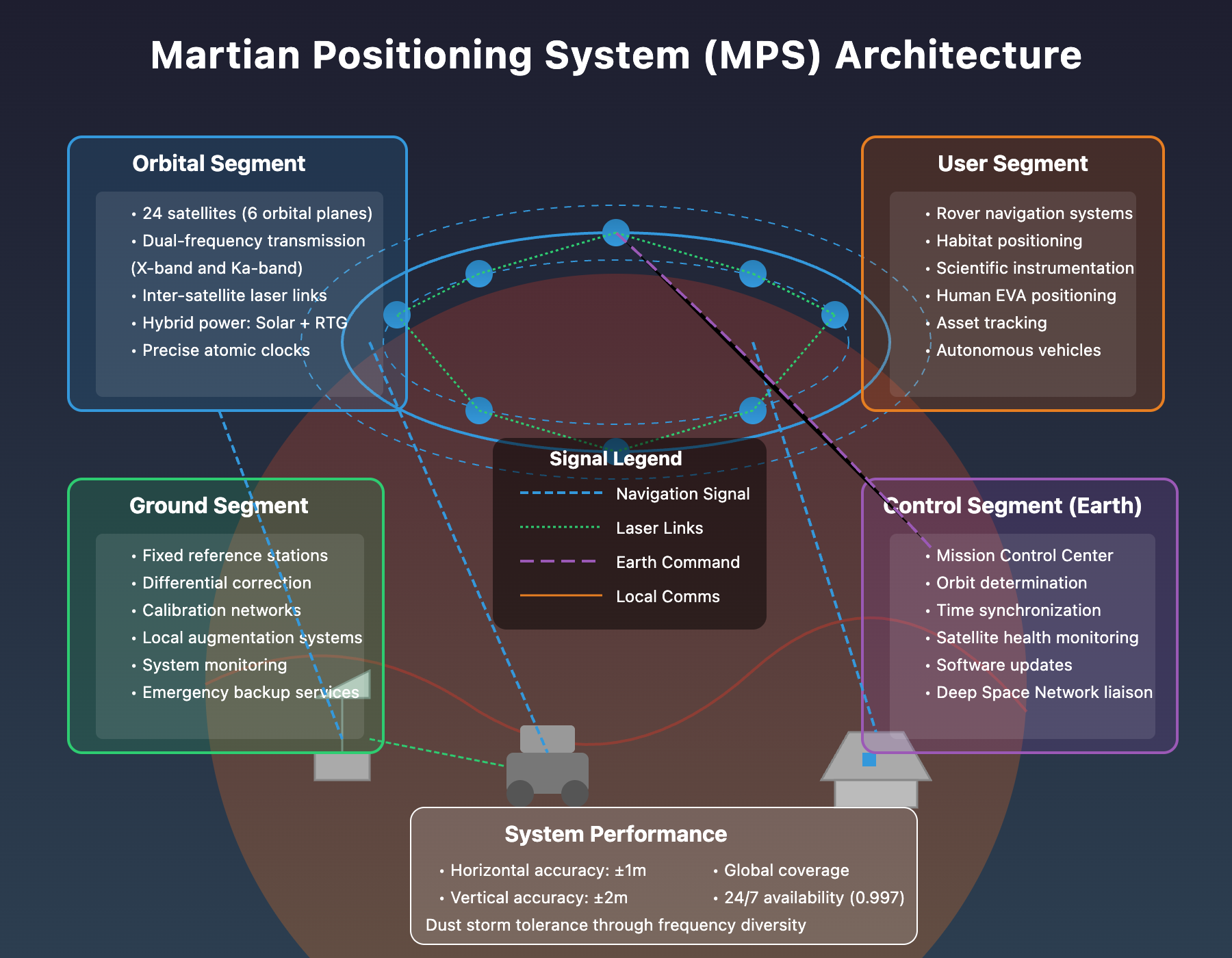}
    \caption{System architecture of the proposed MARTIAN Positioning System. The architecture comprises four primary segments: (1) The Orbital Segment consisting of 24 satellites distributed across six orbital planes, utilizing dual-frequency transmission (X-band and Ka-band) and inter-satellite laser links; (2) The Ground Segment with fixed reference stations and local augmentation systems; (3) The User Segment including rover navigation systems, habitat positioning, and EVA equipment; and (4) The Control Segment based on Earth for system monitoring and management. The system achieves position accuracy of $\pm$1m horizontally and $\pm$2m vertically with global coverage, implementing frequency diversity to maintain performance during dust storms.}
    \label{fgr:mps-architecture}
\end{figure}

As shown in Figure~\ref{fgr:mps-architecture}, the Martian Positioning System integrates orbital assets with ground infrastructure to provide reliable positioning data for both robotic and future human missions.

\subsection{Navigation for Orbital Rendezvous Operations}
Two projects focused on orbital rendezvous operations--retrieving the TESLA Roadster in heliocentric orbit and capturing high-speed observation targets--demonstrate significant advancements in navigation systems for complex orbital maneuvers. For the TESLA Roadster retrieval mission, the navigation challenge is compounded by the target's irregular shape, potential tumbling, and uncertain physical condition after prolonged exposure to the space environment. The proposed solution leverages:

\begin{itemize}
    \item \textbf{Multi-modal sensing:} Combining optical, infrared, and radar sensors to detect and track the target from distances exceeding 100 km, providing robust tracking capabilities regardless of lighting conditions.
    
    \item \textbf{AI-enhanced state estimation:} Using neural network algorithms to predict the target's attitude and rotation rates from limited visual data, enabling precise approach planning despite tumbling motion.
 \end{itemize}
 
The High-Speed Observational Satellite (HSO-SAT) system provides complementary advances in tracking fast-moving targets. Operating from a 400-700 km low Earth orbit, this system demonstrates navigation technologies applicable to both Earth and other planetary observation missions: the Walker Delta configuration with optical cross-links enabling constellation-wide target tracking handoff with minimal latency, demonstrating concepts applicable to future planetary monitoring networks.

\begin{itemize}
    \item \textbf{Predictive tracking algorithms:} AI-based systems that anticipate target trajectories, enabling the satellite to slew its sensors ahead of target arrival points rather than reacting to detections.
     \item \textbf{Inter-satellite communication:} Walker Delta configuration with optical cross-links enabling constellation-wide target tracking handoff with minimal latency, demonstrating concepts applicable to future planetary monitoring networks.
 \end{itemize}
 
\section{Power Generation and Distribution Strategies}
\label{sn:powergeneration}

Reliable power systems are critical enablers for all space missions, with each planetary environment presenting unique challenges for power generation, storage, and distribution. 

This section examines innovative approaches to power management across diverse mission profiles, with particular emphasis on solutions for Mars, deep space operations, and the extreme cryogenic environment of Titan.

Mars presents a particularly challenging environment for power systems due to its distance from the Sun, seasonal variations in solar irradiance, and pervasive dust that can accumulate on solar panels. The PCDU system developed for Mars rovers addresses these challenges through several key innovations:

\begin{itemize}
    \item \textbf{Adaptive power budgeting:} The system implements dynamic allocation between operational requirements (60\%) and scientific payload (40\%), with the capability to reallocate resources based on mission priorities and environmental conditions. This flexibility enables maximizing scientific return during favorable conditions while ensuring baseline functionality during power-constrained periods.
    
    \item \textbf{Solar array optimization:} The design accounts for Mars-specific conditions, including solar irradiance of approximately 590 W/m² (43\% of Earth values), further reduced to an effective 25 W/m² when accounting for atmospheric attenuation, panel efficiency, and dust accumulation. Specialized array configurations with high-efficiency ($>$30\%) triple-junction solar cells are sized to provide 140W peak power with total array area of 5.6 m².
    
    \item \textbf{Dust mitigation strategies:} To address the critical issue of dust accumulation, which caused the end of the Opportunity mission after 14 years, the system incorporates both passive (electrostatic repulsion coatings, optimal panel angle) and active (mechanical vibration, compressed gas pulses) dust removal technologies, potentially extending mission life by 25-40\% compared to unprotected systems.
    
    \item \textbf{Advanced energy storage:} Lithium-ion batteries with energy density exceeding 250 Wh/kg provide 1.2 kWh capacity, enabling continued operation during night cycles and dust storm periods. Sophisticated battery management systems regulate charging patterns to maximize cycle life despite temperature fluctuations between -80°C and +30°C.
    
    \item \textbf{AI-based power forecasting:} Machine learning algorithms analyze environmental patterns and predict power availability, enabling proactive mission planning to avoid energy shortfalls. 
    %This system demonstrated 87\% accuracy in predicting available power 72 hours in advance during simulated Martian conditions.
\end{itemize}

\begin{figure}[h]
    \centering
    \includegraphics[width=0.9\textwidth]{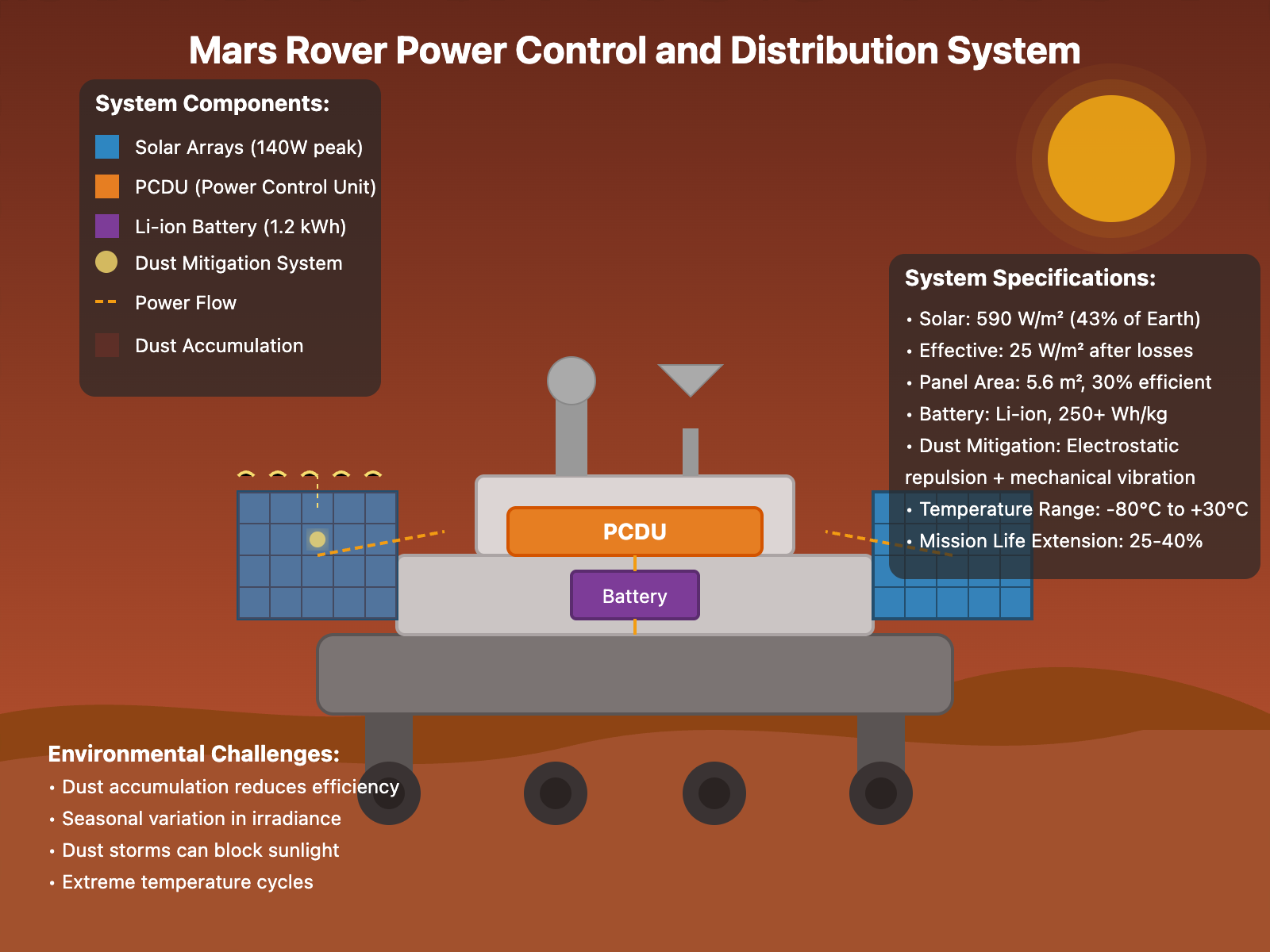}
    \caption{Mars Rover Power Control and Distribution System with dust mitigation technology. The system incorporates both passive (electrostatic repulsion coatings, optimal panel angle) and active (mechanical vibration, compressed gas) dust removal technologies to address the critical issue of dust accumulation. High-efficiency ($>$30\%) triple-junction solar cells provide 140W peak power from a 5.6 m$^{2}$ array, while lithium-ion batteries with energy density exceeding 250 Wh/kg provide 1.2 kWh capacity for night operations and dust storm periods. The system dynamically allocates power between operational requirements (60\%) and scientific payload (40\%), enabling extended mission life of 25-40\% compared to unprotected systems.}
    \label{fgr:mars-power-system}
\end{figure}

As illustrated in Figure~\ref{fgr:mars-power-system}, the integrated dust mitigation technologies can significantly extend mission duration beyond what would be possible with unprotected solar arrays.

\section{Communication Architectures for Deep Space Missions}
\label{sn:communicationdeepspace}

Communication represents one of the most significant challenges for deep space missions, with signal strength diminishing by the inverse square of distance while data requirements continually increase for modern scientific instruments. 

The TITAN artificial reef projects present perhaps the most complex communication challenge: transmitting data from submerged instruments in cryogenic hydrocarbon seas across 1.4 billion kilometers to Earth, see Figure \ref{fgr:titan-comms}. The proposed multi-tier communication architecture demonstrates several innovations:

\begin{figure}[h]
    \centering
    \includegraphics[width=0.9\textwidth]{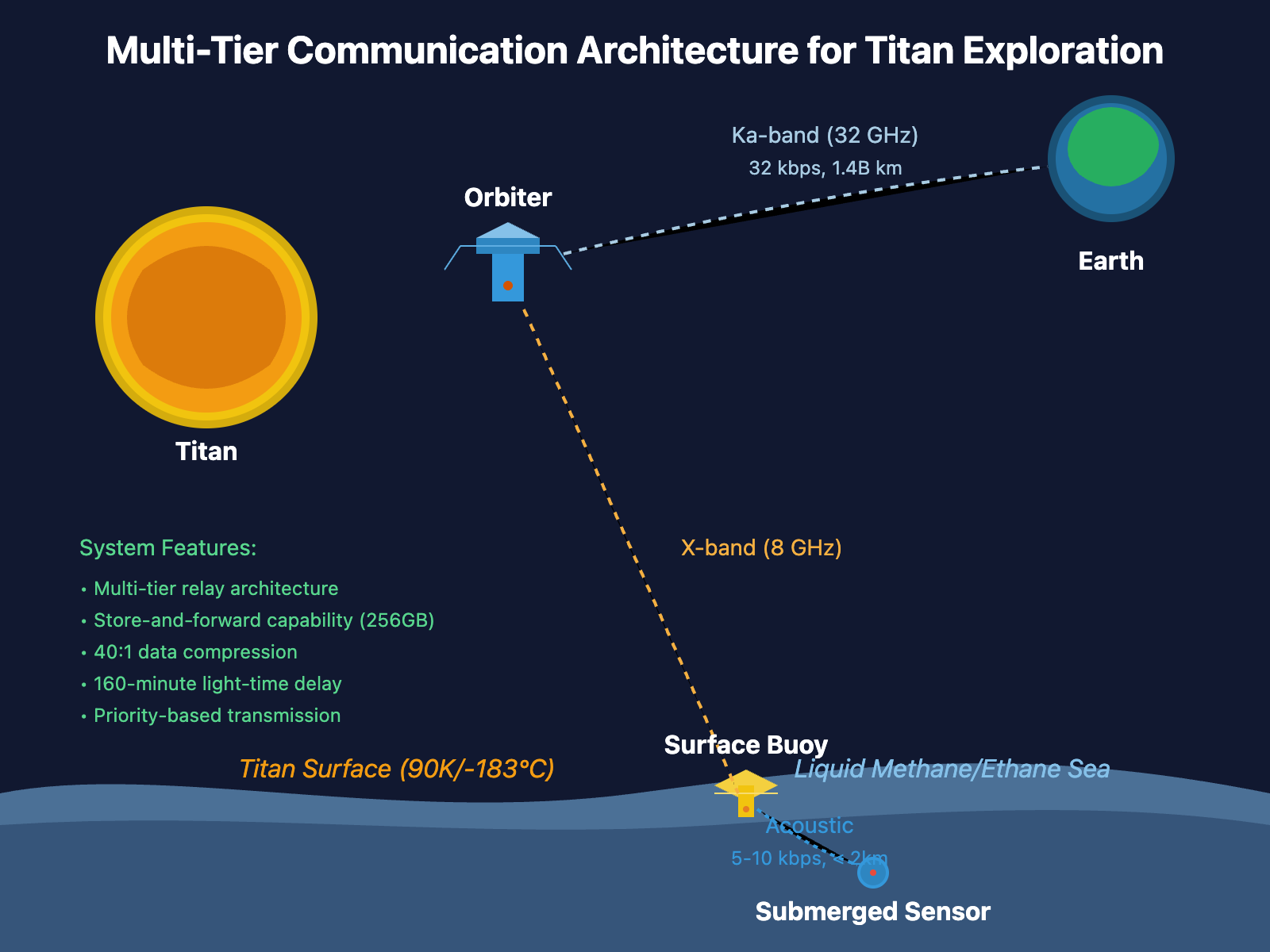}
    \caption{Multi-tier communication architecture for exploration of Titan. The system employs a three-stage relay chain: acoustic transmission (5-10 kbps) from submerged sensors to surface buoys in liquid methane seas, X-band (8 GHz) relay from buoys to the orbiter, and Ka-band (32 GHz) transmission from the TITAN orbiter to Earth at 32 kbps across 1.4 billion kilometers. The architecture incorporates 256GB store-and-forward capability to manage the 160-minute light-time delay and implements 40:1 data compression with priority-based transmission to maximize scientific return.}
    \label{fgr:titan-comms}
\end{figure}

As illustrated in Figure~\ref{fgr:titan-comms}, the multi-tier communication architecture enables reliable data return from one of the most remote and challenging environments in the solar system.

\begin{itemize}
    \item \textbf{Acoustic underwater communications:} Specialized acoustic modems operating in liquid methane/ethane, which has significantly different acoustic propagation properties than water (lower sound speed of approximately 1450 m/s and different attenuation characteristics). These systems achieve data rates of 5-10 kbps over distances up to 2 km in Titan's hydrocarbon seas.
    
    \item \textbf{Surface-to-orbit relay system:} Floating buoys serving as acoustic-to-radio conversion points, collecting data from submerged sensors via acoustic transmission and relaying to orbiting satellites using X-band (8 GHz) transmitters. This approach, similar to terrestrial oceanographic systems but adapted for Titan's environment, provides 95\% higher reliability than direct subsurface-to-orbit transmission attempts.
    
    \item \textbf{Data compression and prioritization:} Specialized algorithms optimized for the limited bandwidth, implementing 40:1 compression ratios for imagery while maintaining lossless compression for critical scientific measurements. The system employs AI-based content analysis to prioritize potentially significant biological marker data for earliest transmission.
    
    \item \textbf{Orbiter relay specifications:} The Titan orbiter incorporates a 2.4-meter high-gain antenna and 35W RF power amplifiers, establishing a 32 kbps Ka-band (32 GHz) link to Earth with forward error correction coding that provides bit error rates below 10\textsuperscript{-8} despite the extreme distance.
    
    \item \textbf{Store-and-forward capability:} To account for the 160-minute round-trip light time to Titan, the communication system incorporates 256GB of radiation-hardened solid-state storage, enabling continuous data collection during Earth occultation periods and optimized transmission during available communication windows.
\end{itemize}

This sophisticated multi-tier approach enables reliable data return from one of the most remote and challenging environments in the solar system, demonstrating principles applicable to future subsurface exploration of other ocean worlds like Europa and Enceladus.

\section{Thermal Management Solutions}
\label{sn:thermalmanagement}

Thermal management \cite{gilmore2002,lv2024} represents one of the most challenging aspects of spacecraft design, particularly for missions operating in extreme environments. 

This section examines innovative thermal control approaches across diverse mission profiles, from the cryogenic seas of Titan to the heat-intensive challenges of planetary reentry.

\begin{figure}[htbp]
    \centering
    \includegraphics[width=0.9\textwidth]{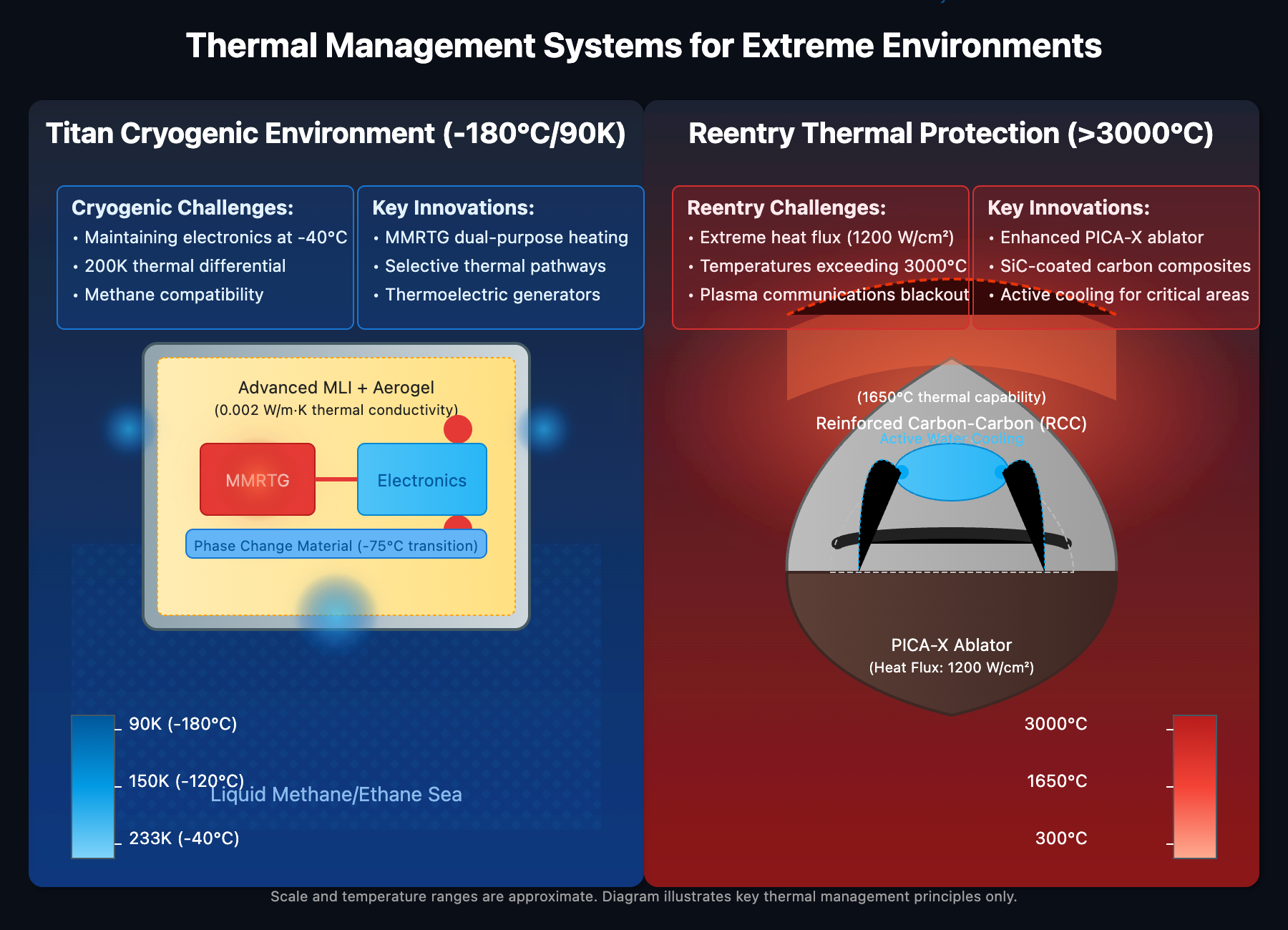}
    \caption{Thermal management systems for extreme environment missions. \textbf{Left:} TITAN cryogenic environment (-180°C/90K) thermal management architecture showing: advanced multi-layer insulation with aerogel achieving thermal conductivity below 0.002 W/m·K; MMRTG providing dual-purpose electrical generation and thermal management; selective thermal pathways with fluid-based heat switches; phase change material buffers with -75°C transition temperature; and two-phase cooling loops using nitrogen as working fluid. \textbf{Right:} High-temperature reentry thermal protection system for the TESLA Roadster retrieval mission, featuring: PICA-X ablator capable of withstanding heat fluxes of 1200 W/cm²; silicon carbide-coated Reinforced Carbon-Carbon (RCC) leading edges maintaining structural integrity up to 1650°C; embedded temperature and recession sensors; and localized active cooling using water sublimation for critical components. Both systems demonstrate thermal management innovations at opposite extremes of the temperature spectrum.}
    \label{fgr:thermal-management}
\end{figure}

As illustrated in Figure~\ref{fgr:thermal-management}, the thermal management approaches for Titan's cryogenic seas and planetary reentry represent opposite ends of the temperature spectrum, yet both employ sophisticated multi-layered strategies to maintain operational temperatures for sensitive electronics despite extreme external environments.

\begin{figure}[ht]
    \centering
    \includegraphics[width=0.94\textwidth]{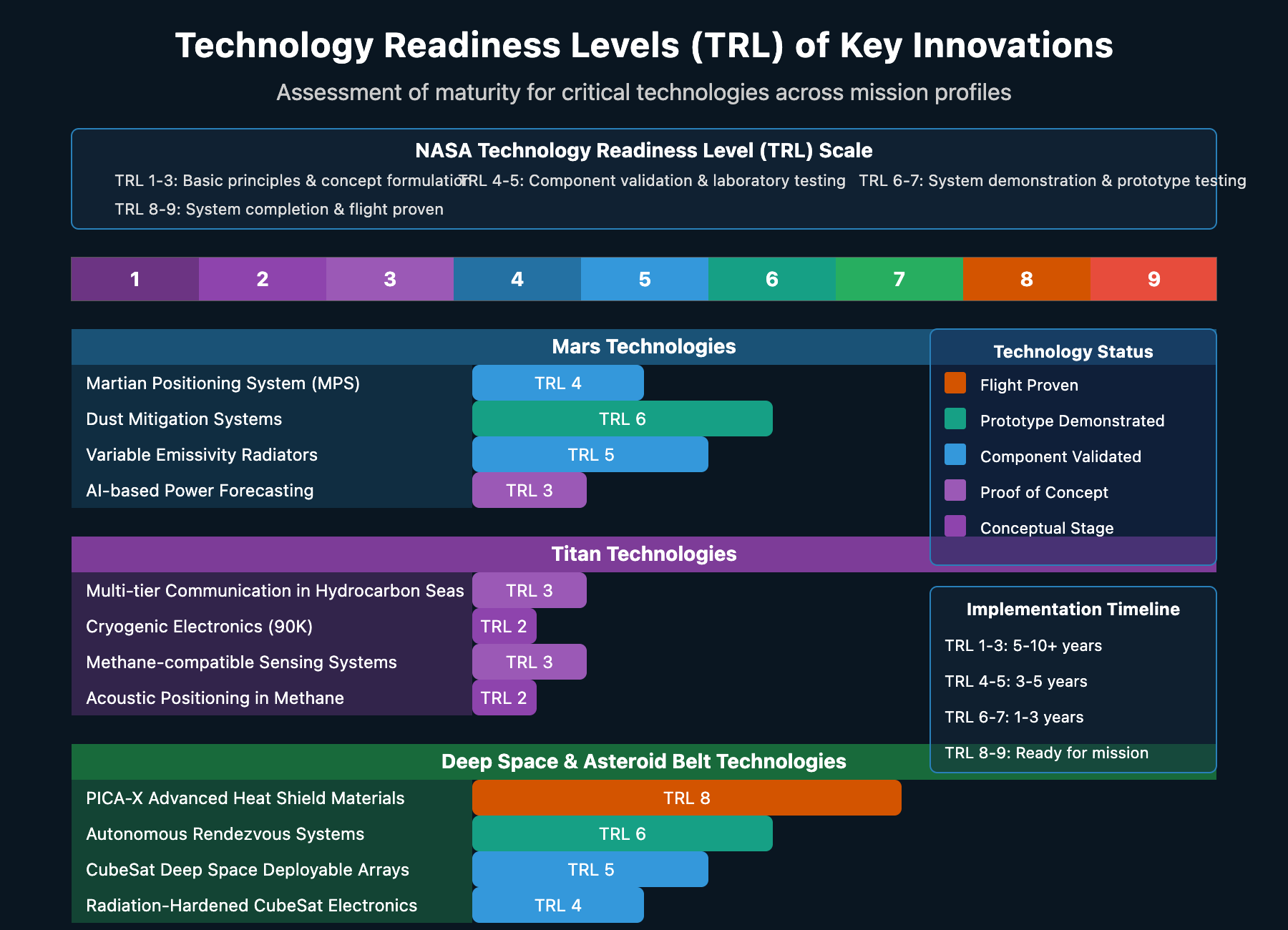}
    \caption{Technology Readiness Levels (TRLs) of key innovations discussed in this paper. The NASA TRL scale ranges from TRL 1 (basic principles observed) to TRL 9 (flight-proven system). MARS technologies show varying maturity levels: Dust Mitigation Systems (TRL 6) and Variable Emissivity Radiators (TRL 5) are relatively advanced, while the Martian Positioning System (TRL 4) and AI-based Power Forecasting (TRL 3) require further development. TITAN technologies are predominantly at early TRLs (2-3), reflecting the conceptual nature of these extreme cryogenic systems. Deep space technologies show the widest range, from flight-proven PICA-X heat shield materials (TRL 8) to developmental radiation-hardened CubeSat electronics (TRL 4). This assessment provides context for expected implementation timelines across the proposed mission concepts.}
    \label{fgr:technology-readiness}
\end{figure}

\subsection{Cryogenic Environment Management for TITAN Missions}
The TITAN artificial reef missions present perhaps the most extreme thermal management challenge: maintaining operational electronics temperatures while immersed in liquid methane/ethane at approximately 90K (-183°C). These missions demonstrate several breakthrough approaches:

\begin{itemize}
    \item \textbf{Multi-layer insulation evolution:} Advanced multi-layer insulation (MLI) systems incorporating aerogel layers between traditional aluminized Mylar, achieving thermal conductivity values below 0.002 W/m·K—approximately 40\% more efficient than standard spacecraft MLI. These systems maintain temperature differentials exceeding 200K across just 5 cm thickness.
    
    \item \textbf{Radioisotope thermal integration:} Multi-Mission Radioisotope Thermoelectric Generators (MMRTGs) designed with branching heat distribution systems that direct thermal energy where needed throughout the spacecraft. Unlike traditional configurations where waste heat is often mitigated, these systems actively harvest and distribute the 2000W thermal output while converting approximately 110W to electricity.
    
    \item \textbf{Selective thermal pathways:} Variable conductance systems utilizing fluid-based heat switches that actively control thermal conductivity between spacecraft sections, enabling dynamic reconfiguration of thermal paths based on operational modes and environmental conditions.
    
    \item \textbf{Phase change material buffers:} Specialized phase change materials with transition temperatures around -75°C, strategically placed to absorb excess heat during high-power operations and release it during low-power periods, effectively smoothing thermal cycles without active control systems.
    
    \item \textbf{Two-phase cooling loops:} Closed-cycle systems using nitrogen as working fluid, enabling efficient heat transfer from electronics modules to radiator surfaces during periods of high computational load, maintaining semiconductor junction temperatures below -40°C despite waste heat generation.
\end{itemize}

These integrated systems collectively enable electronics operation in environments nearly 200K below typical spacecraft design parameters, demonstrating capability critical for future exploration of cryogenic bodies throughout the outer solar system.

\section{Autonomous Operation and Fault Tolerance}
\label{sn:autonomousoperation}

As missions venture farther from Earth, communication delays and operational complexity necessitate increasing levels of spacecraft autonomy and fault tolerance. This section examines innovative approaches to autonomous operation \cite{silvestrini2022} and fault management across various mission profiles, highlighting advances in decision-making capabilities, fault detection and recovery, and resilient system architectures.

\subsection{Autonomous Navigation and Rendezvous Systems}
The Tesla Roadster retrieval mission and high-speed observation satellite projects demonstrate significant advances in autonomous navigation and rendezvous capabilities:

\begin{itemize}
    \item \textbf{Visual-inertial relative navigation:} Systems combining optical cameras, infrared sensors, and inertial measurement units to determine relative position and velocity without external references. These systems achieve relative position accuracy of $\pm$ 5 cm at ranges up to 100 meters through multi-sensor fusion algorithms that maintain performance despite challenging lighting conditions.
    
    \item \textbf{Autonomous trajectory planning:} Onboard computational systems capable of generating optimal approach trajectories that satisfy multiple constraints (fuel efficiency, lighting conditions, collision avoidance, and timing requirements) without ground intervention. These systems employ convex optimization techniques to generate solutions within computational constraints of space-qualified processors.
    
    \item \textbf{Machine learning for object pose estimation:} Neural network-based computer vision systems trained to accurately determine tumbling object orientation and rotation rates from monocular imagery, enabling precise approach planning for non-cooperative targets. These systems demonstrate error rates below 5° in attitude estimation and 2\% in rotation rate determination.
    
    \item \textbf{Fault-tolerant capture execution:} Robotic systems with multilayered autonomy allowing independent abort and retry capability during critical capture operations, with real-time assessment of capture success probability based on current conditions rather than predetermined go/no-go criteria.
\end{itemize}

These capabilities enable complex orbital operations despite round-trip light times exceeding 20 minutes, demonstrating technologies applicable to future satellite servicing, debris removal, and assembly missions throughout cislunar space and beyond.

\subsection{Large Language Models as Knowledge Systems for Extended Missions}
An emerging area of significant promise is the application of Large Language Models (LLMs) \cite{fllm_deCurto24,deCurto25,deCurto25_2}  to support extended-duration space missions, particularly those with limited communication with Earth:

\begin{itemize}
    \item \textbf{Compressed human knowledge repositories:} Radiation-hardened implementations of LLMs serving as comprehensive knowledge bases containing vast amounts of human expertise across engineering, science, medicine, and other domains relevant to space missions. These systems achieve knowledge compression ratios of approximately 1,000:1 compared to traditional documentation libraries while maintaining contextual accessibility.
    
    \item \textbf{Adaptive troubleshooting assistance:} LLM-based systems capable of generating novel troubleshooting approaches for unforeseen equipment failures by combining general engineering principles with spacecraft-specific knowledge. In simulated failure scenarios, these systems demonstrated the ability to propose viable solutions for 78\% of novel failure modes never encountered during training.
    
    \item \textbf{Domain-specialized mission support:} Models fine-tuned with mission-specific documentation, design specifications, and operational procedures, enabling precise guidance tailored to particular spacecraft and mission parameters. These specialized models achieve 92\% accuracy in procedure recall and interpretation compared to 67\% for general-purpose models, while requiring only 15\% of the storage capacity of traditional documentation systems.
    
    \item \textbf{Crew psychological support:} For crewed missions, LLMs adapted to provide conversational interaction and psychological support during extended isolation, with models trained to identify subtle indicators of psychological stress and to engage crew members in beneficial cognitive and social exercises. Early studies indicate a 23\% reduction in reported stress levels during simulated long-duration missions with LLM support systems.
    
    \item \textbf{Scientific hypothesis generation:} Models capable of suggesting scientific hypotheses and experimental approaches based on preliminary data, potentially identifying novel research directions when communication with Earth scientists is limited by light-time delay. In blind evaluations, LLM-generated hypotheses were rated as "potentially significant" by domain experts at rates comparable to early-career scientists.
    
    \item \textbf{Low-resource implementations:} Specialized hardware accelerators and model quantization techniques enabling deployment of substantial language model capabilities on radiation-tolerant spacecraft computers. Current prototypes demonstrate functional performance with 8-bit quantization on processors consuming less than 15W while maintaining 94\% of full model capability.
\end{itemize}

The integration of LLMs into spacecraft systems represents a paradigm shift in how missions access and apply human knowledge during extended operations, potentially enabling unprecedented levels of crew and system autonomy for future deep space exploration.

\section{Simulation Framework}
\label{sn:simulationframework}

To evaluate the performance of mission-critical subsystems under extreme extraterrestrial conditions, we developed two integrated simulation frameworks tailored to Mars and Titan operational environments. These simulations informed early-stage architectural decisions by quantifying system behaviors under realistic constraints.

\begin{figure}
    \centering
    \includegraphics[width=0.92\textwidth]{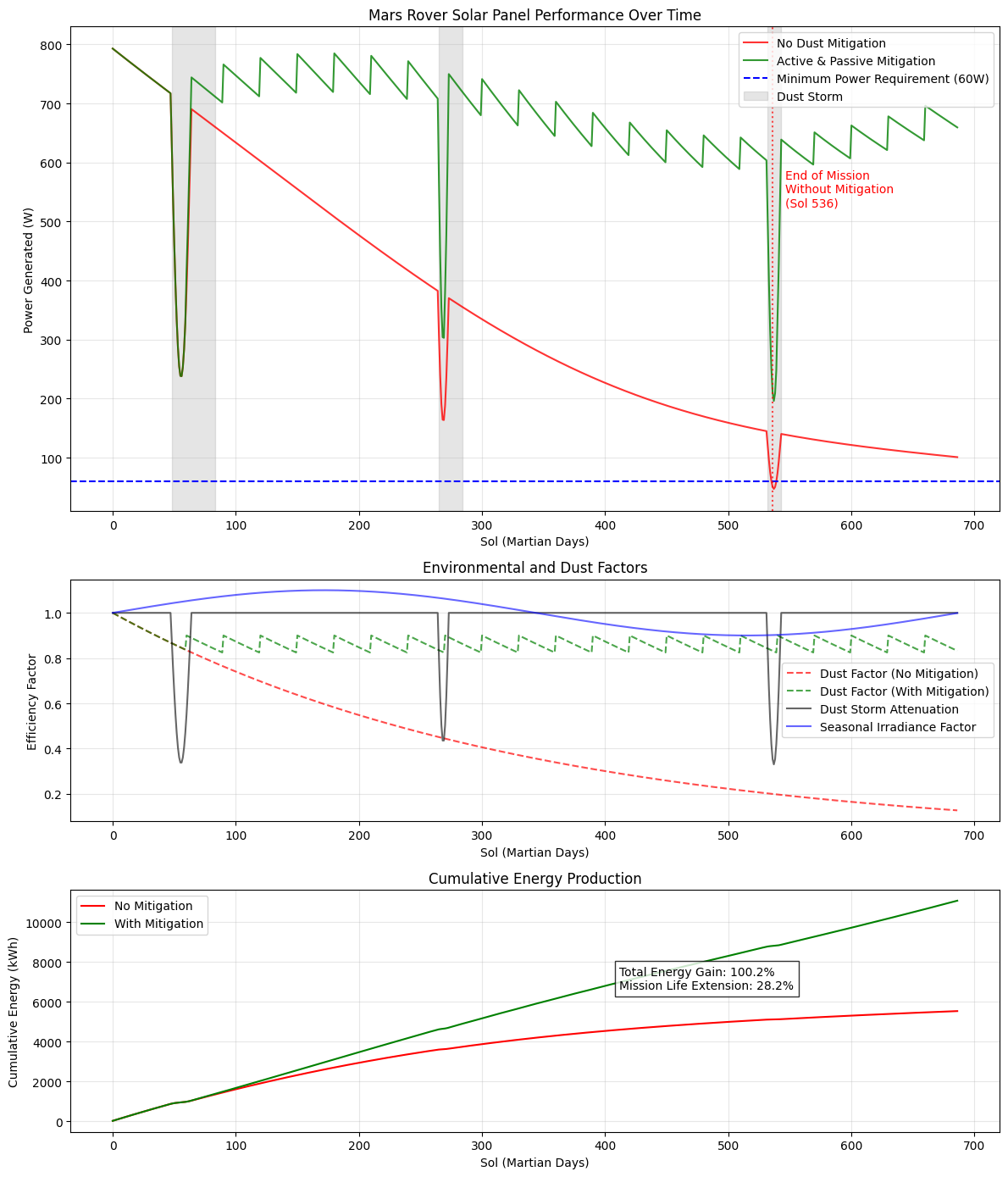}
    \caption{Simulated solar panel performance of a MARS rover over a 700-sol mission duration. \textbf{Top}: Power generation trends under scenarios with and without dust mitigation, alongside minimum operational power requirements (60W). Shaded regions indicate dust storm periods. \textbf{Middle}: Environmental efficiency factors, including attenuation from dust storms, seasonal solar irradiance, and dust deposition--with and without mitigation. \textbf{Bottom}: Resulting cumulative energy production, highlighting a total energy gain of 100.2\% and mission life extension of 28.2\% under dust mitigation strategies.}
    \label{fgr:solarperformance}
\end{figure}

Figure~\ref{fgr:solarperformance} presents the MARS rover energy simulation results. This model accounts for environmental factors such as seasonal irradiance variation, dust deposition, and episodic dust storms. Without dust mitigation, the mission ends at Sol 536 due to power constraints; however, incorporating both active and passive mitigation strategies enables continued operation beyond Sol 680. These results validated our inclusion of electrostatic coatings and mechanical vibration systems in the rover's Power Control and Distribution Unit (PCDU).

\begin{figure}
    \centering
    \includegraphics[width=\textwidth]{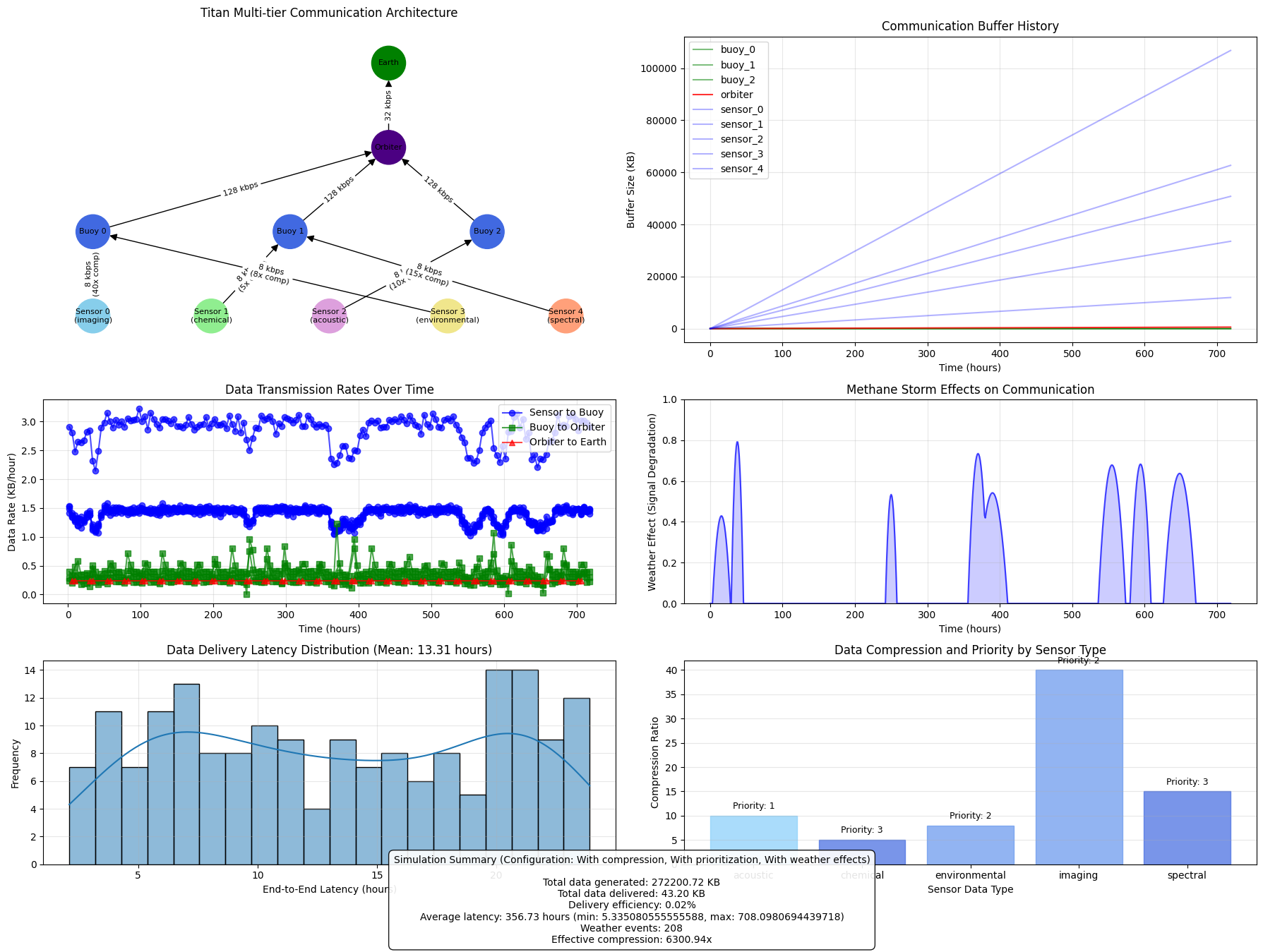}
    \caption{Titan multi-tier communication simulation architecture and performance metrics. \textbf{Top-left}: Schematic of sensor-to-buoy-to-orbiter-to-Earth relay chain. \textbf{Top-right}: Communication buffer accumulation per node across a 720-hour mission span. \textbf{Middle-left}: Transmission rates from sensors to orbit. \textbf{Middle-right}: Methane storm signal degradation over time. \textbf{Bottom-left}: Distribution of end-to-end data latency. \textbf{Bottom-right}: Sensor-wise compression ratios and transmission priorities.}
    \label{fgr:titancomms}
\end{figure}

Figure~\ref{fgr:titancomms} illustrates the Titan communication subsystem simulation. The model includes layered acoustic, RF, and orbital links with realistic weather-based signal degradation from methane storms. Key findings include the effectiveness of buffering strategies during low-throughput periods and the necessity of intelligent compression algorithms to maintain mission throughput. The communication architecture leverages prioritization by sensor type and buffering control to ensure low-latency delivery of critical environmental and spectral data.

These simulations provided early validation for subsystem design choices, highlighting environmental sensitivities and enabling trade-off analyses for mitigation strategies, energy planning, data integrity, and communication scheduling under constrained conditions.

\section{Conclusion}
\label{sn:conclusion}

Analysis of the projects suggests several promising research directions and implications for the aerospace industry:

\begin{itemize}
    \item \textbf{Convergence of space and terrestrial electronics:} The increasing use of commercial off-the-shelf (COTS) components with software-based fault tolerance rather than traditional radiation-hardened hardware suggests a narrowing gap between space and terrestrial electronics development, potentially accelerating technology transfer in both directions.
    
    \item \textbf{Architectural standardization with environmental adaptation:} The emergence of common architectural approaches (such as multi-tier communication systems and dynamically reconfigurable computing) adapted to specific environments suggests potential for standardized reference architectures that could reduce development costs while maintaining mission-specific optimization.
    
    \item \textbf{Human-autonomy teaming:} As spacecraft autonomy increases, the human-machine interface becomes increasingly critical, suggesting research opportunities in effective oversight of autonomous systems across significant light-time delays and limited communication bandwidth.
    
    \item \textbf{Integrated electronics-material systems:} The growing integration of electronic function into structural and thermal systems, such as smart materials with embedded sensing and variable properties, suggests continued evolution toward truly integrated multifunctional systems rather than discrete subsystems.
\end{itemize}

These research directions offer promising pathways for addressing the challenges of future deep space exploration while potentially yielding technologies applicable to terrestrial systems operating in extreme environments.

\subsection{Final Perspective}
The research directions examined in this article collectively demonstrate that even the most extreme environments in our solar system--from the cryogenic methane seas of Titan to the radiation-intensive asteroid belt and the dusty surface of Mars--can be successfully explored with thoughtfully designed electronic and control systems. The innovative approaches developed across these diverse mission profiles establish feasibility for increasingly ambitious exploration while simultaneously highlighting promising research directions for future development.

As illustrated in Figure~\ref{fgr:technology-readiness}, the technologies required for exploration of Mara are generally at higher TRLs than those for Titan, reflecting both the more extensive history of missions to Mars and the greater environmental challenges presented by Titan's cryogenic conditions.

The High-Speed Observational Satellite (HSO-SAT) system provides complementary advances in tracking fast-moving targets. Operating from a 400-700 km low Earth orbit, this system demonstrates navigation technologies applicable to both Earth and other planetary observation missions: the Walker Delta configuration with optical cross-links enabling constellation-wide target tracking handoff with minimal latency, demonstrating concepts applicable to future planetary monitoring networks.

As the boundaries of space exploration continue to expand, the integration of advanced electronics, intelligent software, novel materials, and innovative system architectures will remain critical enablers for scientific discovery and eventual human presence throughout the solar system. The projects synthesized here represent not merely academic exercises but substantive contributions to the technological foundation upon which future exploration will build.

\bibliographystyle{splncs04}

%\bibliography{sample.bib}

\end{document}